\documentclass[fleqn,twoside]{article}
\usepackage{espcrc2}
\usepackage{amssymb}

\usepackage{graphicx}
\usepackage[figuresright]{rotating}

\newcommand{\be}{\begin{equation}}
\newcommand{\ee}{\end{equation}}
\newcommand{\bea}{\begin{eqnarray}}
\newcommand{\eea}{\end{eqnarray}}

\title{Glueball masses in U(1) LGT using the multi-level algorithm}

\author{P. Majumdar\address[MCSD]{Max-Planck-Institut f\"ur Physik,
 F\"ohringer Ring 6, D-80805, M\"unchen}%
        \thanks{e-mail: pushan@mppmu.mpg.de},
        Y. Koma\addressmark\thanks{e-mail: ykoma@mppmu.mpg.de},
        M. Koma\addressmark\thanks{e-mail: mkoma@mppmu.mpg.de}}
       
\begin{document}

\begin{abstract}
The multi-level algorithm allows, at least for pure gauge theories, reliable 
measurement of exponentially small expectation values. The implementation of 
the algorithm depends strongly on the observable one wants to measure. Here 
we report measurement of glueball masses using the multi-level algorithm in 4 
dimensional compact U(1) theory as a case study.
\vspace{1pc}
\end{abstract}

\maketitle

\section{Introduction}
Compact U(1) lattice gauge theory exists in two phases. A confining phase
at strong coupling and a deconfined phase at weak coupling. Monte Carlo 
simulations have established that for the Wilson action, the phase transition 
point corresponds to $\beta=1.011128(11)$ \cite{ALSN}. The order of the 
transition has been debated for a long time. Recent investigations
including finite size scaling analysis suggest a weak first order 
transition \cite{ALSN}.

While accurate measurements of the glueball mass can throw light on the
order of the transition, such measurements are difficult to perform using
conventional methods.
Glueball masses in 4d compact U(1) lattice gauge theory were first measured 
by Berg and Panagiotakopoulos \cite{Oldphoton} using correlations between
Wilson loops. However they could only go to a separation of 2 in lattice 
units for the correlators. In fact except for values of the coupling 
close to the phase transition point \cite{Stack} (where the glueball is lighter), 
measurements have been carried out only for small temporal lengths of 
the correlators.

Recently L\"uscher and Weisz have proposed an exponential noise reduction
method which exploits the local nature of the action and existence of a 
positive definite transfer matrix \cite{LW1}. In this work we apply this
method, which lets us go to large temporal separations, to glueball correlators
and obtain results which have very little contamination from higher states.
A similar study was also carried out in \cite{Meyer}.

\section{Glueball correlators in the multi-level scheme}

Glueball correlators generically look like 
\be\label{corr}
\langle C(t,t_0) \rangle_{\rm conn} =\langle {\cal O}(t){\cal O}(t_0) \rangle
-\langle {\cal O}(t)\rangle\langle{\cal O}(t_0) \rangle .
\ee
The zero momentum scalar and axial-vector correlators are given by
the operators ${\cal O}_1$ and ${\cal O}_2$ such that 
\bea
{\cal O}_1(t)&\equiv&\sum_{\vec x}\sum_{ij=1,2,3}{\bf Re}(P_{ij}({\vec x},t)) \\
{\cal O}_2(t)&\equiv&\sum_{\vec x}{\bf Im}(P_{ij}({\vec x},t)),
\eea
where $P_{ij}$ is the plaquette
in the $ij$ plane. Glueballs with definite momenta ${\vec k}$ are created by 
\be
{\cal O}({\vec k},t)=\sum_{\vec{x}} {\cal O}(\vec{x},t) e^{i\vec{k}\cdot\vec{x}}.
\ee
As a function of the time separation $\Delta t=t-t_0$ the glueball
correlator is expected to behave like
\begin{equation}
\langle C(t,t_0) \rangle_{\rm conn}\approx \alpha \left [
e^{-m\Delta t}+e^{-m(N_t-\Delta t)} \right ],
\end{equation}
where $N_t$ is the extent of the lattice in the time direction.
Fitting the measured correlator to this form, one can obtain the effective mass 
$m$ of the glueball. For zero momentum, the effective mass is equal to the rest
mass while for non-zero momentum, one has to take into account the momentum contribution
($m_k$) to the effective mass to obtain the rest mass.

In the multi-level scheme, $\langle {\cal O}(t){\cal O}(t_0) \rangle$ is estimated 
by $\langle [{\cal O}(t)][{\cal O}(t_0)] \rangle$ where $[\cdots]$ denotes an 
intermediate level of averaging called the sub-lattice average \cite{LW1}.
This scheme requires partial updates of the lattice.
In contrast to a full update where all links are updated, a partial update affects 
links only in a part of the lattice with the boundary of this part held fixed.
The sub-lattice averaging reduces the fluctuation of each individual operator 
${\cal O}$ to a great extent.
This is efficient because the small expectation values are now generated by 
multiplication rather than fine cancellation of positive
and negative values of the same order. 

As long as $\langle{\cal O}\rangle=0$, which is true for the axial-vector
correlator, this procedure works quite well. However in the scalar channel,
where $\langle{\cal O}\rangle\neq 0$ much of
this advantage is lost as each expectation value is
a number ${\mathrm O}(1)$, but the connected part is several orders of magnitude smaller
than the full correlator.
To get around this problem, one can take the derivative of the correlator in the scalar
channel to remove
the VEV of the plaquettes. So let us now take the derivative of
the correlator at both $t$ and $t_0$. \footnote{In principle one derivative is enough,
 but in practice we found that the efficiency of the algorithm is higher for the double
derivative compared to the single one.}
Taking $\partial_t$ to be the forward derivative and $\partial_{t_0}^*$ to be the
backward derivative on the lattice, we get,
\bea\label{dcorr1}
\partial_t\partial_{t_0}^* \langle C(t,t_0) \rangle \approx
-\alpha \left [ e^{-m(t-t_0)}(1-e^{-m})^2 \right . && \nonumber \\
\left . + e^{-m(N_t-(t-t_0))}(e^m-1)^2\right ]. &&
\eea

Now we are far better suited to apply the multi-level scheme. We can now measure
\bea\label{dcorr2}
\partial_t\partial_{t_0}^* \langle C(t,t_0) \rangle = \langle \sum_{{\vec x},i,j}  
\left [ P_{ij}({\vec x},t\!+\!1)-P_{ij}({\vec x},t) \right ] && \nonumber \\
\times\sum_{{\vec x},i,j} \left [ P_{ij}({\vec x},t_0)-P_{ij}({\vec x},t_0\!-\!1) 
\right ] \rangle. && 
\eea

The derivatives were estimated in the sub-lattice updates by taking the difference
of the value of the operator on the updated slice with the value of the operator
on the boundary. As shown in Figure 1, to get the forward derivative at $t$, we used the
fixed boundary at $(t+1)$ and for the backward derivative, the boundary at $(t-1)$.  
To get the correlators one has to use two such slices (e.g. $t$ and $t_0$ in
Figure 1).
In practice we held every alternate layer of spatial links fixed and estimated
the correlators for various temporal separations in the same sweep. 
The only drawback at the moment seems to be the fact that we
have to consider a minimum separation of two in the temporal direction.

\begin{figure}[t]
\includegraphics[angle=0,width=12pc]{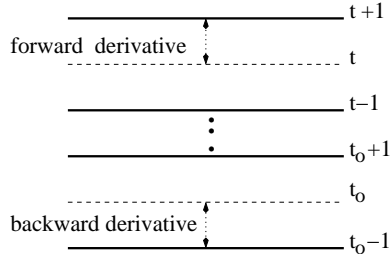}
\vspace*{-4mm}\caption{ Evaluation of the derivative of the glueball correlator. The thick 
lines are time slices held fixed during the sub-lattice averaging.}
\end{figure}

\begin{figure}[htb]\label{opt}
\includegraphics[angle=0,width=16pc]{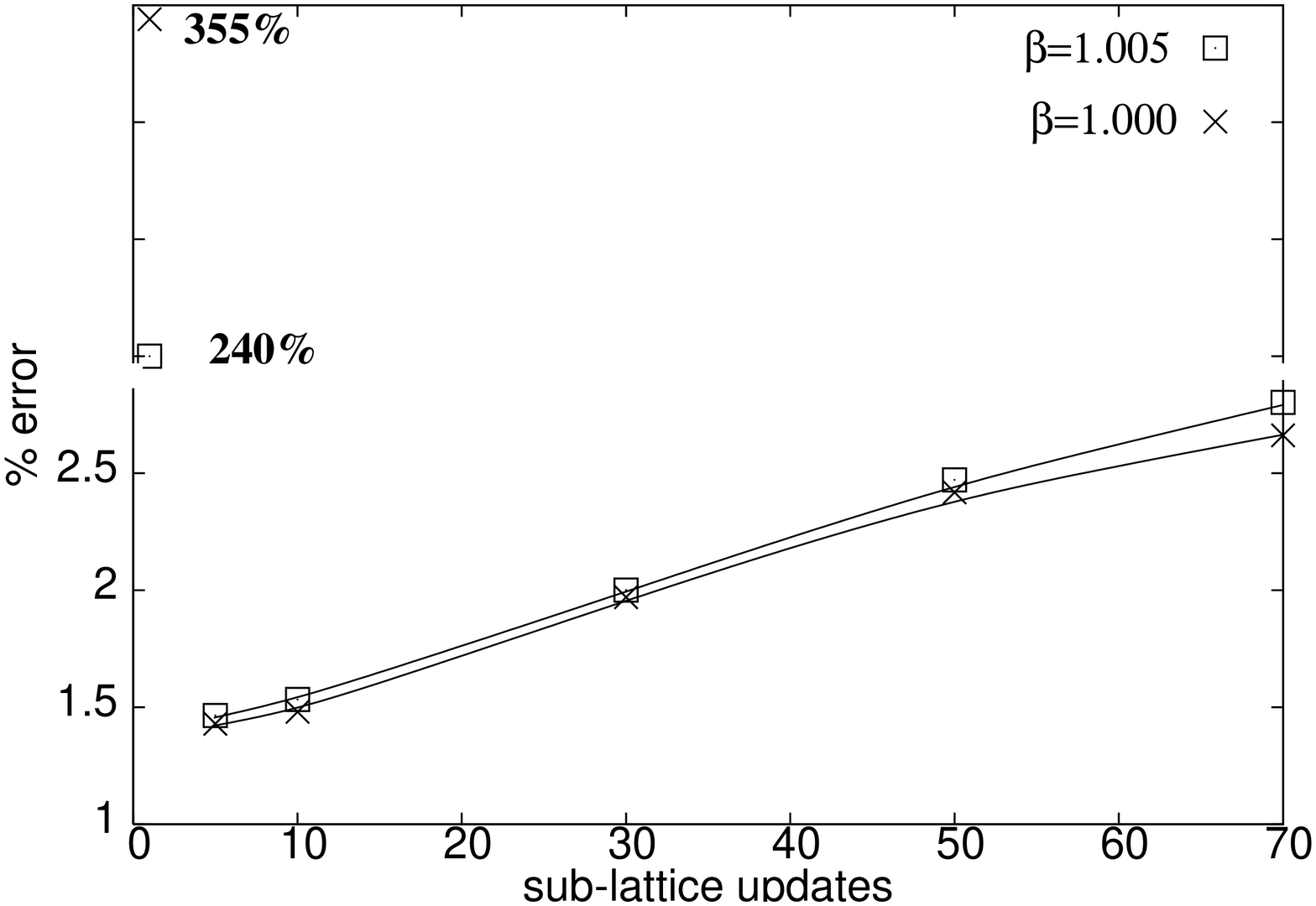}
\includegraphics[angle=0,width=16pc]{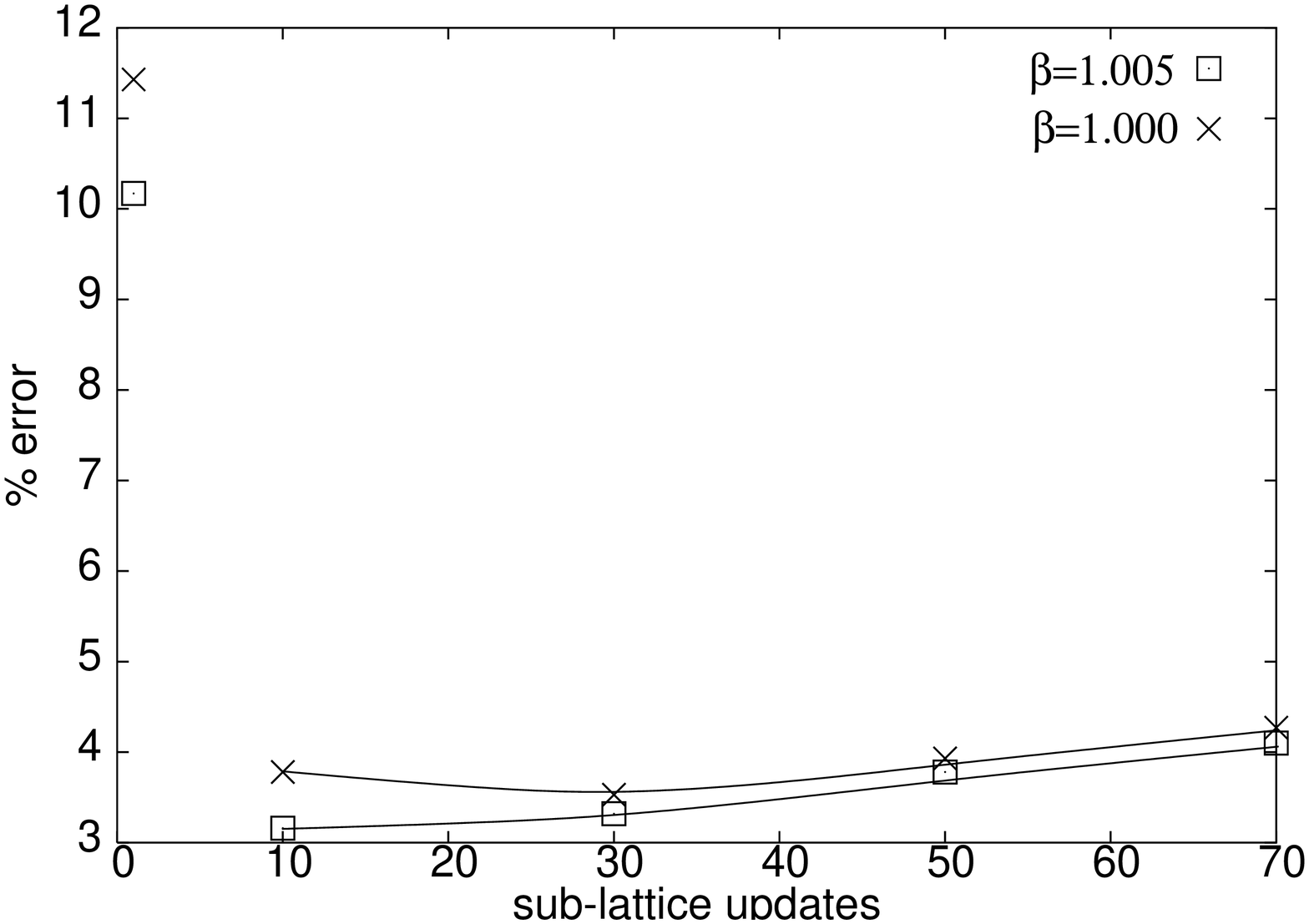}
\vspace*{-5mm}\caption{\%error on the scalar (top) and axial-vector (bottom)
 correlator at $\Delta t=2$ after
a 10hr. run on a 1.5GHz AMD Athlon PC. $12^4$ lattice.} 
\end{figure}

The number of sub-lattice updates is an optimization
parameter of the algorithm that has to be tuned for efficient performance.
This is a function of $\beta$. In the range of $\beta$ we looked at, we found
that 10 to 50 sub-lattice updates were sufficient.
To compare this procedure with the naive algorithm we measured the percentage
error on the correlators at a value of $(t-t_0)$ where both methods gave non-zero
signals. In a similar amount of computer time, the multi-level algorithm produced
errors which were about two orders of magnitude lower than the naive method.
In Figure \ref{opt} we show the \%error for a given CPU time as a function of 
sub-lattice updates for the scalar and axial-vector channels.

\section{Results}

In Tables \ref{scalarmass} and \ref{vectormass} we present our results
for the zero momentum scalar and axial-vector glueball masses. 

The (${\vec k}\neq0$) axial-vector correlator is sensitive 
to a correlation between two 1-forms. In the deconfining region we expect this part to yield 
information about the photon. 
Indeed the effective masses from this correlator are very small in the deconfined phase
and do not show any variation with $\beta$.
To obtain the rest mass, we computed $m^2-m^2_k$ assuming
the free field dispersion relation
$m^2_k=\sum_i(2-2\cos k_i)$. Within statistical errors the rest mass
turns out to be zero and this we believe is strong evidence
for the photon. For a more elaborate discussion 
of our results see \cite{PYM}. 

\begin{figure}[t]\label{masses}
\includegraphics[angle=0,width=16pc]{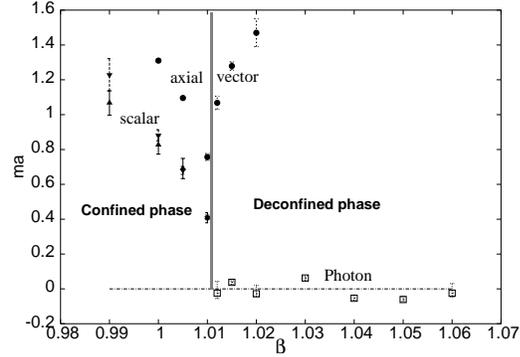}
\vspace*{-4mm}\caption{Consolidated plot of glueball masses.
$\blacktriangle$ and $\blacktriangledown$ :
scalar channel (${\vec k}=0$) from $\partial\partial^* \langle C
\rangle$ and $\partial^*\partial \langle C \rangle $.
$\bullet$ : axial-vector channel (${\vec k}=0$).
$\square$ : photon from the axial-vector correlator.}
\end{figure}

In Figure \ref{masses}, we present all the masses
along with the region where the phase transition is expected to take place.

\begin{table}[b]
\vspace*{-15mm}\caption{\label{scalarmass} Zero momentum scalar glueball masses}
\begin{tabular}{ccc}
\hline
$\beta$ & $\partial_t^*\partial_{t_0}\langle C(t,t_0)\rangle $& 
$\partial_t\partial_{t_0}^*\langle C(t,t_0)\rangle $\\
\hline
0.990 & 1.195 (55) & 1.085 (50) \\
1.000 & 0.875 (41) & 0.812 (37) \\
1.005 & 0.682 (29) & 0.693 (29) \\
1.010 & 0.405 (22) & 0.410 (21) \\
\hline
\end{tabular}
\\ \\
\caption{\label{vectormass} Zero-momentum axial-vector glueball masses}
\begin{tabular}{c|c|c|c}
\multicolumn{4}{l}{Confining regime} \\
\hline
$\beta$ & 1.000 & 1.005 & 1.010 \\
\hline
$ ma $   & 1.31 (1) & 1.096 (9) & 0.757 (20) \\
\hline 
\multicolumn{4}{l}{}\\
\multicolumn{4}{l}{Deconfining regime} \\
\hline
$\beta$ & 1.012 & 1.015 & 1.020 \\
\hline
$ma$ & 1.068 (38) & 1.279 (24) & 1.47 (8) \\
\hline
\end{tabular}
\end{table}

\end{document}